\documentclass[aip,pop]{revtex4-1}

\usepackage{amsmath}
\usepackage{graphicx}
\usepackage{mathptmx}

\newcommand{\dee}{\textnormal{d}}

\newcommand{\mb}[1]{\mathbf{#1}}

\begin{document}

\title{Plasmas generated by ultra-violet light rather than electron impact}
\date{\today}
\author{R.\ N.\ Franklin}
\email{raoulnf1935@gmail.com}
\affiliation{Department of Astronomy and Physics, The Open University, Milton Keynes, MK7 6AA, UK}
\author{J.\ E.\ Allen}
\affiliation{University College, University of Oxford, Oxford, OX1 4BH, UK; OCIAM, Mathematical Institute, University of Oxford, Oxford, OX2 6GG, UK}
\affiliation{Blackett Laboratory, Imperial College London, Prince Consort Road, London, SW7 2BW, UK}
\author{D.\ M.\ Thomas}
\affiliation{Blackett Laboratory, Imperial College London, Prince Consort Road, London, SW7 2BW, UK}
\author{M.\ S.\ Benilov}
\affiliation{Departamento de Fisica, CCCEE, Universidade de Madeira, Largo do Municipio, 9000, Funchal, Portugal}

\begin{abstract}
We analyze, in both plane and cylindrical geometries, a collisionless plasma consisting of an inner region where generation occurs by UV illumination, and an un-illuminated outer region with no generation. Ions generated in the inner region flow outwards through the outer region and into a wall. We solve for this system's steady state, first in the quasi-neutral regime (where the Debye length $\lambda_D$ vanishes and analytic solutions exist) and then in the general case, which we solve numerically. In the general case a double layer forms where the illuminated and un-illuminated regions meet, and an approximately quasi-neutral plasma connects the double layer to the wall sheath; in plane geometry the ions coast through the quasi-neutral section at slightly more than the Bohm speed $c_s$. The system, although simple, therefore has two novel features: a double layer that does not require counter-streaming ions and electrons, and a quasi-neutral plasma where ions travel in straight lines with at least the Bohm speed. We close with a pr\'{e}cis of our asymptotic solutions of this system, and suggest how our theoretical conclusions might be extended and tested in the laboratory.
\end{abstract}

\maketitle

\section{Introduction}

This paper is concerned to give a description of plasmas where the generation is by photo-ionization rather than electron impact.

An experimental and theoretical treatment of the situation we have in mind was given by Johnson, Cooke, and Allen\cite{Johnson78} who in cylindrical geometry passed the output of a mercury discharge through a vessel containing mercury vapour, both vessels joined by a quartz window so passing UV.  In their situation the ionization was associative caused by the excited states 6$^3$P$_0$ and 6P$^3_1$ interacting according to the scheme Hg($6^3$P$_0$) + Hg($6^3$P$_1$) $\rightarrow$ Hg$_2^+$ + e, to produce a molecular ion Hg$_2^+$ in the $\Sigma$ state. The relevant term diagram was given in Forrest and Franklin,\cite{Forrest69} with earlier measurements having been carried out by Tan and von Engel.\cite{Tan68}

Here we will treat cases where the radiation is sufficiently energetic that ionization occurs directly and the scheme h$\nu$ + A $\rightarrow$ A$^+$ + e applies, that is, direct photo-ionization.

For simplicity we will assume the plasma to be collisionless and we will cover both plane and cylindrical geometries.

We begin with the plasma approximation, where results not very different from the positive column are expected.
We then consider in detail partial illumination with a maximum in the centre where the most interesting case is that of a sharp cutoff and it turns out that it is necessary to introduce Poisson's equation to obtain closure.
We then summarize the use of matched asymptotic approximations to explore the double layer that forms in the partially illuminated case in plane geometry.
Finally we point up ways in which this work could be extended and applied.

\section{Theoretical model}

\begin{figure}
\includegraphics[width=0.4\textwidth]{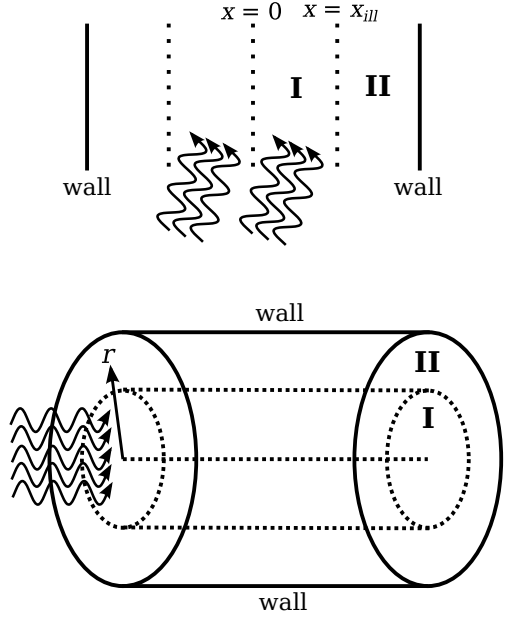}
\caption{\label{fig:sys} The system to be analyzed, in plane geometry (upper) and cylindrical geometry (lower). In both geometries the system comprises two regions of plasma: the central, uniformly illuminated region I and the outer, un-illuminated region II.}
\end{figure}

We show in Figure \ref{fig:sys} a schematic diagram of the experimental situation we envisage when the plasma is generated by end illumination. Other geometries are possible.

Four equations describe the system at equilibrium. Two are steady-state fluid equations, namely the continuity equation
\begin{equation}
\nabla \cdot (n_i \mb{v}) = G
\label{eq:unnormcont}
\end{equation}
and the ion momentum equation
\begin{equation}
Mn_i (\mb{v} \cdot \nabla) \mb{v} + M \mb{v} G = - n_i e \nabla V
\end{equation}
while the other two equations are Poisson's equation
\begin{equation}
\nabla^2 V = \frac{e}{\varepsilon_0} (n_e - n_i)
\end{equation}
and the Boltzmann relation
\begin{equation}
n_e = n_0 \, \exp \left( \frac{eV}{k T_e} \right)
\end{equation}
where $n_i$ and $n_e$ are the ion and electron densities respectively, $\mb{v}$ is the ion velocity, $G$ the generation rate, $M$ the ion mass, $V$ the electric potential, $T_e$ the electron temperature, and $n_0$ the electron density at the system's central axis. The equations are the same in both regions, but $G$ is zero in the un-illuminated region II. We use $G$ rather than $Z$ in (\ref{eq:unnormcont}) because the equation with electron impact ionization reads $\nabla \cdot (n_i \mb{v}) = Zn$ and is not simply integrable (although it can be solved analytically in the quasi-neutral case).

Introducing the normalized quantities $\mb{U} \equiv \mb{v}/c_s$, $N \equiv n/n_0$, $X \equiv x/L_{ion} \equiv x / (n_0 c_s / G)$, and $\Phi = -eV/(k_B T_e)$, we rewrite the above equations in normalized form:
\begin{equation}
\tilde{\nabla} \cdot (N_i \mb{U}) = 1
\label{eq:normcont}
\end{equation}
\begin{equation}
\frac{\mb{U}}{N_i} + (\mb{U} \cdot \tilde{\nabla}) \mb{U} 
= \tilde{\nabla} \Phi
\label{eq:normmom}
\end{equation}
\begin{equation}
\tilde{\nabla}^2 \Phi = \frac{N_i - N_e}{\Lambda^2}
\end{equation}
\begin{equation}
N_e = \exp( -\Phi )
\label{eq:normboltz}
\end{equation}
in region I, where $\Lambda \equiv \lambda_D / L_{ion}$ and $\tilde{\nabla}$ is a normalized gradient operator (the spatial derivative when distance is normalized by $L_{ion}$). In Region II (\ref{eq:normcont}) reduces to $\tilde{\nabla} \cdot (N_i \mb{U}) = 0$ and (\ref{eq:normmom}) reduces to ($\mb{U} \cdot \tilde{\nabla}) \mb{U} = \tilde{\nabla} \Phi$.

\section{Quasi-neutral solution in plane geometry}

\label{S:plaqn}

We first consider region I in plane geometry under the plasma approximation, which leads to a quasi-neutral solution. Exploiting symmetry we reduce (\ref{eq:normcont})--(\ref{eq:normboltz}) to scalar equations, and using the plasma approximation $(N_i = N_e = N)$, Poisson's equation is dispensed with. There are then three equations to solve:
\begin{equation}
\quad \frac{\dee (NU)}{\dee X} = 1
\label{eq:cont1}
\end{equation}
\begin{equation}
\frac{U}{N} + U \frac{\dee U}{\dee X} = \frac{\dee \Phi}{\dee X}
\label{eq:ionmo1}
\end{equation}
\begin{equation} N = \exp(- \Phi) \label{eq:br1} \end{equation}

It is readily found that the equations are singular where $U=1$, i.e.\ the Bohm criterion applies. Applying the boundary condition that $U=0$ at $X=0$, (\ref{eq:cont1}) implies $NU = X$ and we obtain the analytic solution
\begin{equation}
X = \frac{U}{1+U^2}, \  N = \frac{1}{1+U^2}, \  \Phi = \ln \left( 1+U^2 \right)
\end{equation}
which implies $U = 1$, $X = 0.5$, $N = 0.5$, $\Phi = \ln \, 2$ at the singularity. The corresponding values for electron impact ionization, previously given by Franklin,\cite{Franklin76} are $X = 0.571$, $N = 0.5$, $\Phi = \ln 2 = 0.693$.

In region II the three equations to solve are more trivial because there is no generation: $NU$ is constant, $U \, \dee U / \dee X = \dee \Phi / \dee X$, and $N = \exp(-\Phi)$. Eliminating $N$ and $\Phi$,
\begin{equation}
U \frac{\dee U}{\dee X} = \frac{1}{U} \frac{\dee U}{\dee X}
\end{equation}
and this apparently has two possible solutions, $\dee U / \dee X = 0$ or $U = 1$. In either case the ions are `coasting', and the only satisfactory solution is to have the ions coasting at the Bohm speed, but then the density is constant and region II is apparently limitless in extent.

Thus this outer region is an unusual type of plasma. It is generationless, it has constant density and, as at a sheath's edge, its ions travel at the Bohm speed, but unlike a sheath it is not just a few Debye lengths long. Although unusual, it does satisfy Langmuir's definition of a plasma as a ``region containing balanced charges of ions and electrons''.\cite{Langmuir28} We might call it a quasi-plasma, but perhaps quiescent plasma is a better description. At a certain time historically there were whole conferences given over to quiescent plasmas, the last being that in Elsinore in 1971,\cite{Elsinore71} which emphasized the difference then perceived between fusion plasmas and low temperature, low pressure plasmas.

\section{Finite Debye length solution in plane geometry}

We now reintroduce Poisson's equation
\begin{equation}
\frac{\dee^2 \Phi}{\dee X^2} = \frac{N_i - N_e}{\Lambda^2}
\end{equation}
thus abandoning the plasma approximation. Then the problem is tractable only by employing computational methods.

\begin{figure*}
\includegraphics[width=0.99\textwidth]{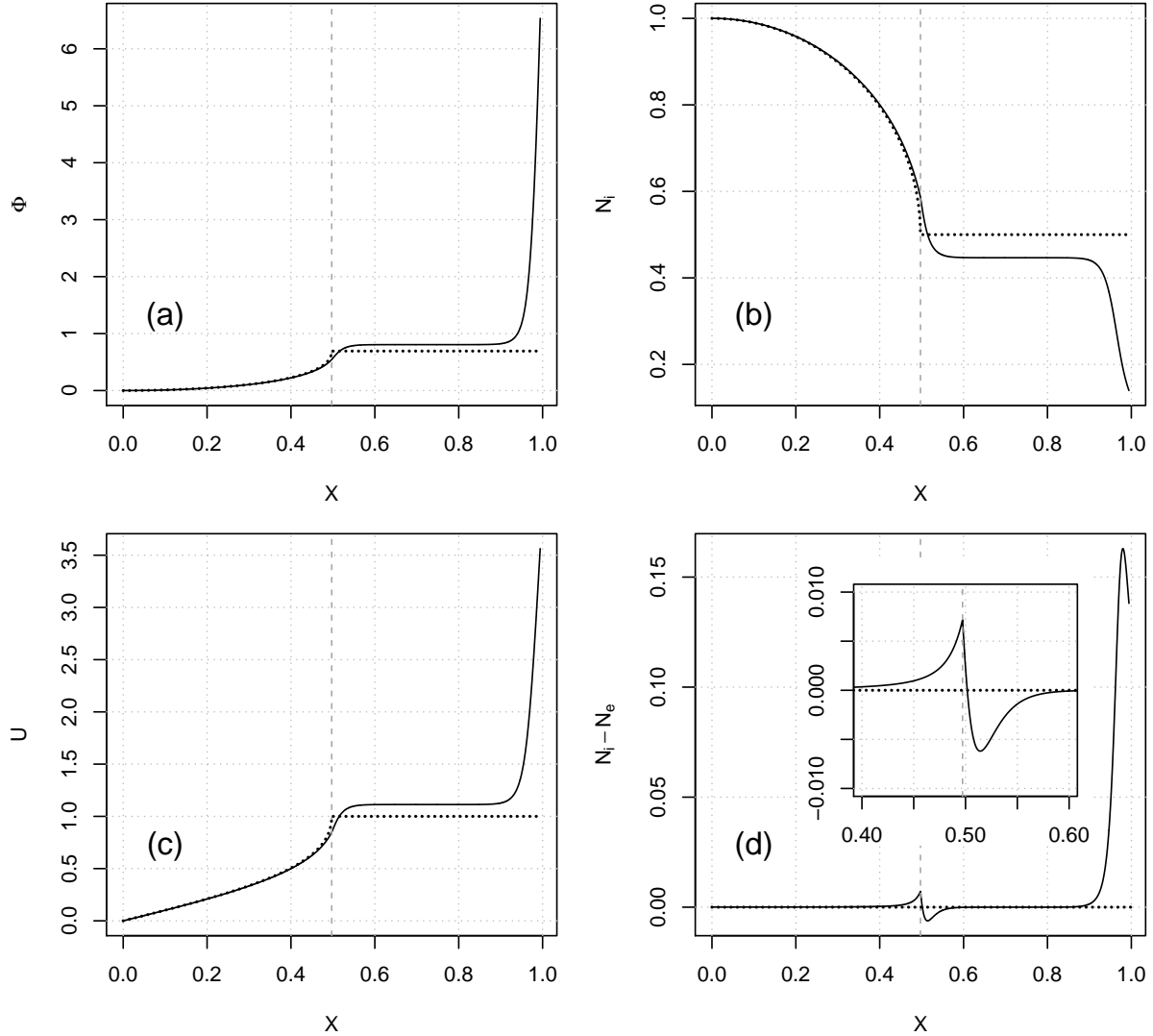}
\caption{\label{fig:pla} Solutions of the model in plane geometry. Dotted curves are the quasi-neutral solution and solid curves the numerical solution for Hg$_2^+$ ions ($\lambda_D / x_{ill} = 1/100$ and $M/m_e = 736744$; the wall sheath depends quantitatively on $M/m_e$). The normalized variables plotted here are (a) electric potential; (b) ion density; (c) ion speed; and (d) net space charge density, demonstrating the existence of a double layer (inset in (d)).}
\end{figure*}

A natural consequence of introducing Poisson is that the computational output gives the space charge difference directly, and one finds that at the boundary between the two regions of plasma, one active and the other quiescent, a double layer forms.

Such structures have been known since the 1970s between plasmas with different characteristics of density or electron temperature due particularly to the work of Andrews and Allen.\cite{Andrews71}  The size of the double layer is a few Debye lengths. It is commonly supposed that double layers are a feature of situations only with counter-streaming ions and electrons, but that is not so in this case.

Thus, having a `plasma' in which the densities are not exactly equal, one can integrate to a point where the electron random flux equals the ion directed flux, and this is the standard condition to determine the position of the wall.

We show in Figure \ref{fig:pla} results for a specific case ($M / m_e = 736744$, $\Lambda = 0.00497$) that gives all of the relevant variables and also shows the existence of a double layer. Varying the parameters produces results having the same features.

\section{Quasi-neutral solution in cylindrical geometry}

\label{S:cylqn}

In cylindrical geometry we have the continuity equation
\begin{equation}
\frac{\dee (NUR)}{\dee R} = 2R
\end{equation}
and the ion momentum equation
\begin{equation}
\frac{U}{N} + U \frac{\dee U}{\dee R} = \frac{\dee \Phi}{\dee R}
\end{equation}
with the Boltzmann relation remaining the same. The normalization is the same but our spatial coordinate is now the radial distance $R \equiv Gr / (2 n_0 c_s)$.

Singularity occurs again where $U=1$. In region I we have
\begin{equation}
R = \frac{U}{(1 + 2U^2)^{3/4}}, \ N = \frac{R}{U}, \ \Phi = -\ln N
\end{equation}
giving values at the boundary between regions I and II of $U = 1$, $N = 0.439$, and $\Phi = 0.824$. These compare with the electron impact values of $U = 1$, $N = 0.420$, and $\Phi = 0.869$.

We next consider the un-illuminated region II, in the light of what we have learned from the plane case, using the plasma approximation, and there we now have $\dee (NUR) / \dee R = 0$, i.e.\ $NUR$ constant. With $R$ increasing, $NU$ is necessarily decreasing, and on physical grounds $N$ is decreasing and $U$ increasing. Going through the algebra of the equations we find
$(U - 1/U) \, \dee U / \dee R = 1/R$. $U$ and $R$ are both positive definite functions of $R$ and thus any solution requires $U \ge 1/U$ and so $U \ge 1$. It is worth reminding ourselves that Bohm's original statement\cite{Bohm49} of what became known as his criterion was $Mv^2 \ge k_B T_e$ (in other words, $U \ge 1$) for a sheath to form. Once again we have two different `plasmas' adjoining, but in this case there is continuity of all of the variables and their derivatives, and the fundamental wall-sheath requirement obtains. This is because the equation above allows the possibility that simultaneously $\dee U / \dee R \rightarrow \infty$ and $(U - 1/U) \rightarrow 0$, at the illumination edge.

\section{Finite Debye length solution in cylindrical geometry}

\begin{figure*}
\includegraphics[width=0.99\textwidth]{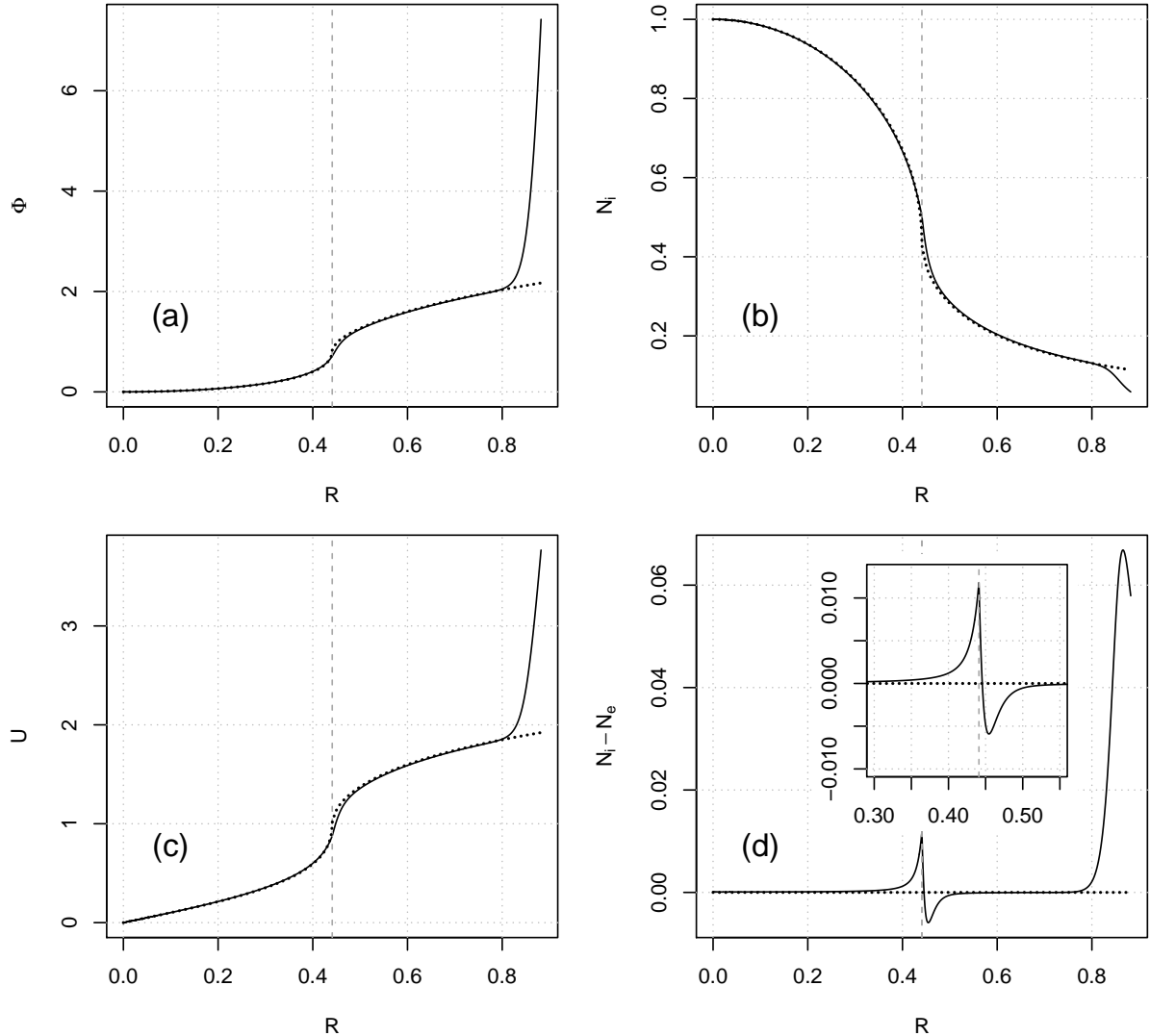}
\caption{\label{fig:cyl} Solutions of the model in cylindrical geometry. Dotted curves are the quasi-neutral solution and solid curves the numerical solution for Hg$_2^+$ ions ($\lambda_D / r_{ill} = 0.00859$ and $M / m_e = 736744$, where $r_{ill}$ is the unnormalized radius of the illuminated region I). The same variables are plotted here as in figure \ref{fig:pla}. An inset detail of (d) again highlights a double layer.}
\end{figure*}

As in plane geometry one has to resort to computation to provide a physically reliable solution to the finite Debye length case, but that we have done. The relevant equations are again (\ref{eq:normcont})--(\ref{eq:normboltz}), and as in the plane case one can flatten them into scalar functions of $R$ by exploiting symmetry.
We show as Figure \ref{fig:cyl} the results for $M/m_e = 736744$ and $\Lambda = 0.00758$. Agan there is a double layer where regions I and II meet.

\section{Joining illuminated and un-illuminated plasmas using matched asymptotic approximations}

The first treatment of transition from an active plasma generated by electron impact to a collisionless near-wall space-charge sheath by means of matched asymptotic expansions in both the fluid model and the free-fall model was given by Franklin and Ockendon.\cite{Franklin70}  It was found that there was a need for an intermediate layer or transition layer intermediate in dimension between the Debye length and the plasma dimension, adjacent to the wall.

The method of matched asymptotic approximations involves expanding the variables in a series in a small parameter, and here we take the illuminated and un-illuminated regions to be of the same extent to reduce the number of parameters involved and designate $\epsilon = (\lambda_D / x_{ill})^2$.
The matching process requires much careful manipulation and so we omit the details here, and one has to proceed order by order.
A first-order solution describing the illuminated plasma, the double layer, and the un-illuminated plasma represents the quasi-neutral solution found in Section \ref{S:plaqn} above for the plane case and in Section \ref{S:cylqn} for the cylindrical case. We will elaborate the full second-order solution in a subsequent paper, but in summary: we find that the double layer scales as $\epsilon^{2/5}$; the second-order terms in expansions of the quantitites $U$, $N_i$, $N_e$, and $\Phi$ are of the order of $\epsilon^{2/5}$ in the illuminated region, of the order of $\epsilon^{1/5}$ in the double layer, and of the orders of, respectively, $\epsilon^{1/5}$ and $\epsilon^{2/5}$ in the un-illuminated region in the plane and cylindrical cases.
It is possible to go to higher orders than the second but the algebra becomes increasingly tedious, and it is more expedient to use computational methods.

\section{Extensions and conclusions}

It is readily possible to introduce collisions in the above theoretical treatments, but this can only be done computationally and in the process the double layer will be smoothed out, typically for $\lambda_i < \lambda_D$ or $c_s/\nu_i < \lambda_D$.

We have given here only results for a sharp cut-off of the illuminated region, but in our earlier exploratory work we looked at situations where the illumination had a profile, either Gaussian or approximating to a Heaviside function. Those situations gave results consistent with the description we have given but that then the double layer attenuated and died in a characteristic parametric manner.

Any experimental test of our results would require a high power UV source, a quite large cylindrical quartz vessel containing the target gas, and ideally non-intrusive diagnostics of the two plasma regions, e.g.\ Laser Induced Fluorescence.  There would need to be a `dump' for the UV that passed through the target gas, but the theory could be readily modified if the attenuation length of the UV was greater than the vessel length, for then the variation with the beam path would be gradual, and not react back on the plasma overall.

Producing the situation in which plane geometry prevailed would be more complicated, but not impossible, and would generate the quiescent uniform plasma we have revealed.

We believe that we have found a new and interesting situation in what we have called the partially illuminated case when there is an active plasma adjacent to a quiescent one before a wall sheath develops.  The physical situation is essentially identical to that which obtains when an external electron beam passes through a plasma, only the parameters are symbolically different. But a recent example producing similar solutions was presented at ICPIG: an ion beam passing through a plasma independently generated, with the same parent gas.\cite{Shumilin13}

\section{Acknowledgement}

Work at Universidade da Madeira was supported by FCT of Portugal through the project PTDC/FIS-PLA/2708/2012.

\end{document}